\documentclass[12pt,preprint]{aastex}

\begin{document}

\title{Consequences of Disk Scale Height on LISA Confusion Noise from Close White Dwarf Binaries}
\author{M. Benacquista}
\affil{Dept. of Sciences, Montana State University-Billings, Billings, MT 59101}
\and
\author{K. Holley-Bockelmann}
\affil{Dept. of Astronomy and the Center for Gravitational Wave Physics, Pennsylvania State University, University Park, PA 16802}

\begin{abstract}
Gravitational radiation from the Galactic population of close white dwarf binaries (CWDBs) is expected to produce a confusion-limited signal at the lower end of the sensitivity band of the Laser Interferometer Space Antenna (LISA). The canonical scale height of the disk population has been taken to be 90 pc for most studies of the nature of this confusion-limited signal. This estimate is probably too low, and the consequences of a more realistic scale height are investigated with a model of the LISA signal due to populations of close white dwarf binaries with different scale heights. If the local space density of CWDBs is held constant, increasing the scale height results in both an increase in the overall strength of the confusion-limited signal as well as in increase in the frequency at which the signals become individually resolvable. If the total number of binaries is held constant, increasing the scale height results in a reduction of the number of expected bright signals above the confusion-limited signal at low frequencies. We introduce an estimator for comparing this transition frequency that takes into account the signal spreading at higher frequencies.
\end{abstract}


\section{Introduction}\label{intro}
The Laser Interferometer Space Antenna (LISA) is a planned space-based gravitational radiation detector that will have a sensitivity band in the range of $10^{-5}$ to $10^{-1}$ Hz. Many of the target sources for LISA will be supermassive and intermediate mass black hole inspirals, as well as extreme mass ratio inspirals of stellar mass compact objects into these more massive black holes. At the lower end of the sensitivity band, the primary noise source will not be instrumental, but rather will come from the accumulated signal from millions of close white dwarf binaries (CWDBs) in the Galaxy. If we interpret the signal as many overlapping sources, this signal is expected to dominate the instrumental noise above about 0.2 mHz. At higher frequencies, the signal will start to separate into individually resolvable sources as the number density of ultra-compact binaries falls with increasing frequency. In most studies of the sensitivity and science capabilities of LISA, this confusion limited signal is modeled as an additional gaussian and stationary noise source in the frequency range between $0.1 - 3$ mHz~\citep[e.g.:][]{cutler98,barack04a,barack04b}.

The overall shape and placement of the confusion-limited noise source is usually based on the original work of~\citet{hbw90} who calculated the expected signal from a number of different binary systems in the Milky Way. The expected signal due to close white dwarf binaries was shown to dominate all other Galactic sources between $0.1$ and $3$ mHz. Their calculation of the expected population of close white dwarf binaries was ostensibly based on a surface density star formation rate, which is independent of the scale height of any Galactic spatial distribution model. However, they then compared the local space density of their model with observation and concluded that a factor of 10 reduction of the total number of binaries would bring their model more in line with observations. This amounts to a {\it de facto} calibration of the total number of binaries by local space density. The calculation of the local space density is dependent upon their Galactic spatial distribution model~\citep{hbw90}:
\begin{equation}
\rho({\bf r}) = \frac{N}{4\pi R_0^2 z_0}e^{-R/R_0}e^{-\left|z\right|/z},
\end{equation}
which is a double exponential with a radial scale $R_0 = 3.5~{\rm kpc}$ and a disk scale height $z = 90$ pc. The choice of $z_0 = 90~{\rm pc}$ follows from the fact that the mass of the close white dwarf binary progenitors are relatively massive, lying in a critical range between $1.12-5.6~{\rm M_{\odot}}$~\citep{webbink84}. It is this reduced population that is used to generate the white dwarf binary background in the {\it LISA Sensitivity Curve Generator}~\citep{larson03}.

More recent simulations of the Galactic white dwarf binary population have used a number of different spatial distributions~\citep{edlund05,nelemans03,nelemans01b}. The most recent simulation of~\citet{edlund05} used the probability distribution:
\begin{equation}
\label{sechsquare}
{\cal P}(R,z) = \frac{1}{4\pi R_0^2z_0}e^{-R/R_0}{\rm sech}^2\left(z/z_0\right),
\end{equation}
where $R_0$ and $z_0$ are the radial scale and scale height, respectively.
Others have included a bulge by effectively doubling the star formation rate in the inner regions of the Galaxy and adding a bulge density of~\citep{nelemans03}:
\begin{equation}
\rho_{\rm bulge} \propto e^{-\left(r/\left(0.5~{\rm kpc}\right)\right)^2},
\end{equation}
where $r = \sqrt{x^2+y^2+z^2}$ in kpc. The Galactic white dwarf model used to generate the confusion noise of~\citet{barack04a,barack04b} is based on the population synthesis of~\citet{nelemans01b}. Through an error in the radial probability distribution function, the spatial distribution actually used was of the form:
\begin{equation}
\label{cuspyrho}
\rho({\bf r}) = \frac{N}{4\pi R R_0 z_0}e^{-R/R_0}{\rm sech}^2(z/z_0),
\end{equation}
which unintentionally does a reasonable job of simulating a bulge~\citep{nelemans03}. In all three cases above, the radial scale was $R_0 = 2.5~{\rm kpc}$ and the scale height was $z_0 = 200$ pc.

It is well-known fact different classes of stars have different scale heights~\citep{gilmore83, kuijken89, kent91, marsakov95, ojha96, reed00, siebert03}, and that for main sequence stars, the more massive ones live in thinner disks~\citep[e.g.:][]{miller79}. While it is true that the current crop of high mass main sequence stars are found in a thin disk, close white dwarf binaries are evolved systems, many of which are much older than the ${\cal O}(10^7)$ years implied by the main sequence lifetimes of the most massive close white dwarf binary progenitors. Any evolved population is known to exhibit a large scale height that is proportional to its vertical velocity dispersion~\citep{nelson02, mihalas81, wielen77, allen73}. There is some debate as to {\it why} old stellar populations live in thicker disks. Perhaps the thickness is simply an artifact of the conditions of the primordial galaxy, when stars were formed from gas clumps within a galaxy potential that was still collapsing to form the disk~\citep{eggen62}, or were easily scattered by an environment rich in galaxy mergers~\citep{steinmetz02}. On the other hand, interactions with potential fluctuations from molecular clouds and other small-scale perturbations may heat an initially cold stellar population, creating an ever thicker disk as it ages and experiences more encounters~\citep{schroder03}.

Whatever the reason, it is clear that the vertical scale height for white dwarfs is rather thick---current estimates place $z_0$ between 240 and 500 pc~\citep{nelson02}. In light of these estimates, we have chosen scale heights of 90 and 500 pc to investigate the consequences of a larger scale height on the nature of the expected gravitational wave signal from close white dwarf binaries in the Galaxy. To minimize the effect of using different population synthesis models, we compare these two scale heights using our own population model. We anticipate similar behavior if one were to change the scale height of a different population model. In order to compare the resulting gravitational wave signals, we introduce a tool for characterizing the separation of strong signals in the frequency domain.

\section{Population Model}\label{popmodel}

The population model we use to generate the different populations of the close white dwarf binaries is modified slightly from the population model described in~\citet{benacquista04}. In this model, the binary types at birth are assigned a probability based on the population synthesis of~\citet{nelemans01a}, assigned an age based on an assumed constant birth rate, and then placed in the galaxy based on probability distributions in cylindrical coordinates $R$ and $z$ given by:
\begin{eqnarray}
P(R)~dR = \frac{R~dR}{R_0^2}e^{-R/R_0} \\
P(z)~dz = \frac{dz}{2 z_0}{\rm sech}^2(z/z_0)
\end{eqnarray}
where $R_0$ is the radial scale length, and $z_0$ is the scale height. In order to investigate the effects of different scale heights, we have chosen to keep a fixed value of radial scale $R_0 = 2500~{\rm pc}$ and have varied only the scale height and the total number of binaries. The space density of binaries, $\rho({\bf r})$, is found from the total number, $N$, and the probability distributions by:
\begin{equation}
\rho({\bf r}) = N{\cal P}(R,z),
\end{equation}
where ${\cal P}$ is given by Equation~\ref{sechsquare}. We assume that the solar neighborhood is located at $R_e = 8500~{\rm pc}$ and $z_e = 0~{\rm pc}$ and define the local space density to be $\rho_e = \rho(R_e,z_e)$. We then obtain $\rho_e = \left(4.25\times 10^{-10}~{\rm pc}^{-2}\right) N/z_0$ for the local space density. For our baseline thin disk model (Thin), we choose $z_0 = 90~{\rm pc}$ and $N = 3\times 10^6$, so that $\rho_e = 1.4\times 10^{-5}~{\rm pc}^{-3}$. We have chosen this value to most closely mimic the reduced population used by~\citet{hbw90}. To obtain the same local space density using the distribution of~\citet{nelemans01b} (Eq~\ref{cuspyrho}), one would need to take $N = 2.2 \times 10^7$ $\left(1.0 \times 10^7\right)$ for a scale height of 200 pc (90 pc). Assuming that roughly 10\% of white dwarfs are in binary systems and about 10\% of these are close white dwarf binaries with orbital period less than $2\times 10^4$ s, this is about a factor of three below a conservative estimate of the expected local white dwarf space density of $4 \times 10^{-3}~{\rm pc}^{-3}$~\citep{knox99}. For our larger scale height model (Thick A), we choose $z_0 = 500~{\rm pc}$ and $N = 1.67\times 10^7$. This model has the same local space density as our baseline model. To investigate the effect of the increased total number in Thick A, we have also introduced a model with the same total number as the Thin model, $N = 3\times 10^6$, but a scale height of $z_0 = 500~{\rm pc}$ (Thick B). These three models are summarized in Table~\ref{models}.
\clearpage
\begin{deluxetable}{lrrr}
\tablecaption{Disk Population Model Parameters\label{models}}
\tablehead{\colhead{Model} & \colhead{$z_0$ (pc)} & \colhead{$N$ ($\times 10^6$)} & \colhead{$\rho_e$ ($\times 10^{-6}~{\rm pc}^{-3}$)}}
\startdata
Thin & 90 & 3 & 14 \\
Thick A & 500 & 16.7 & 14 \\
Thick B & 500 & 3 & 2.6 \\
\enddata
\end{deluxetable}
\clearpage

\begin{figure}
\plotone{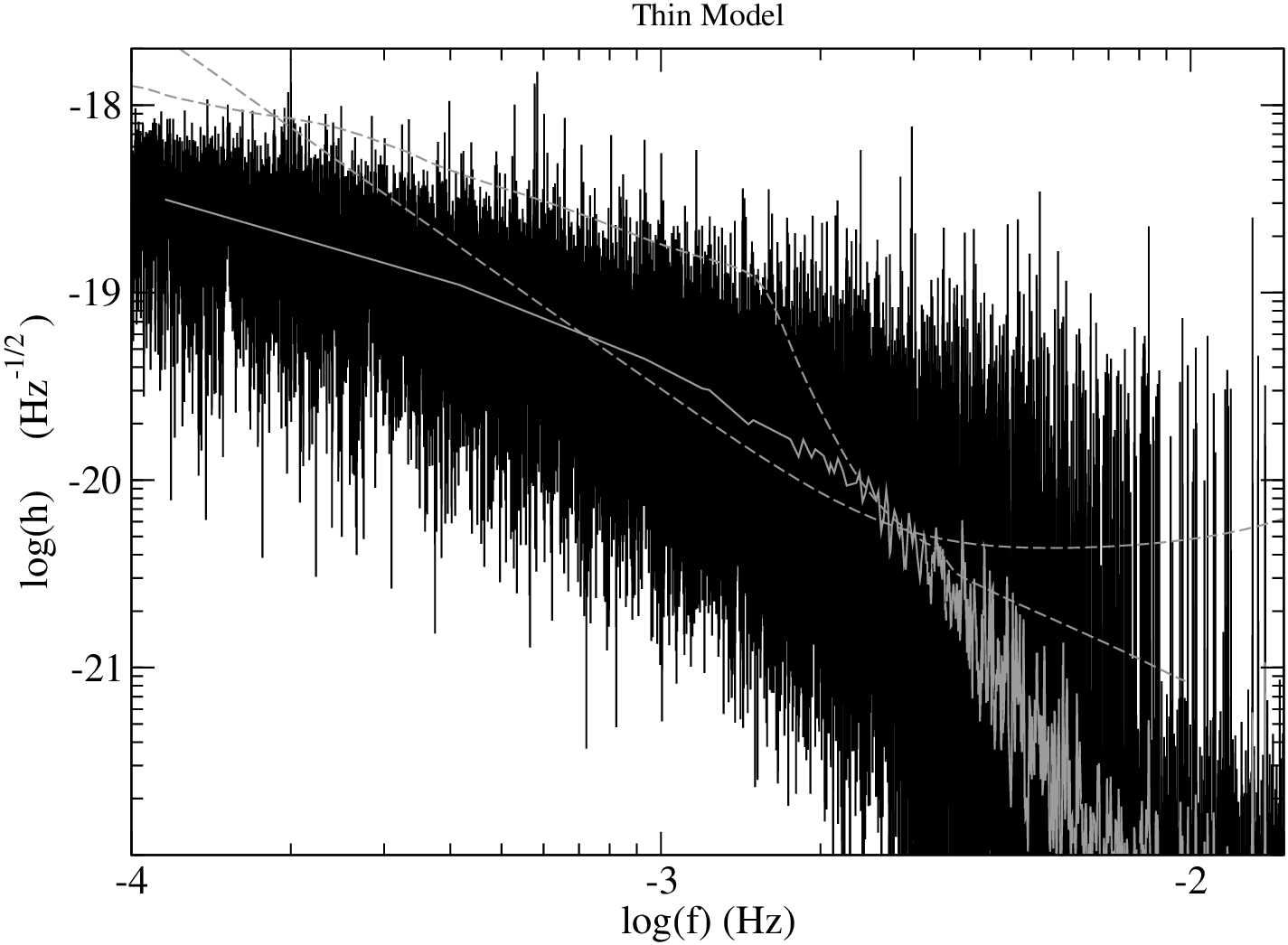}
\caption{Power Spectral Density (black) for the Thin model with a running median over 1000 bins (gray). The expected noise level and standard~\citet{hils97} confusion limited noise from the LISA Sensitivity Curve Generator are shown as the dashed light gray curves. The noise and standard confusion limited noise curves have been multiplied by $\sqrt{3/20}$ to transform them from the barycenter frame to the detector frame.\label{thin}}
\end{figure}

\begin{figure}
\plotone{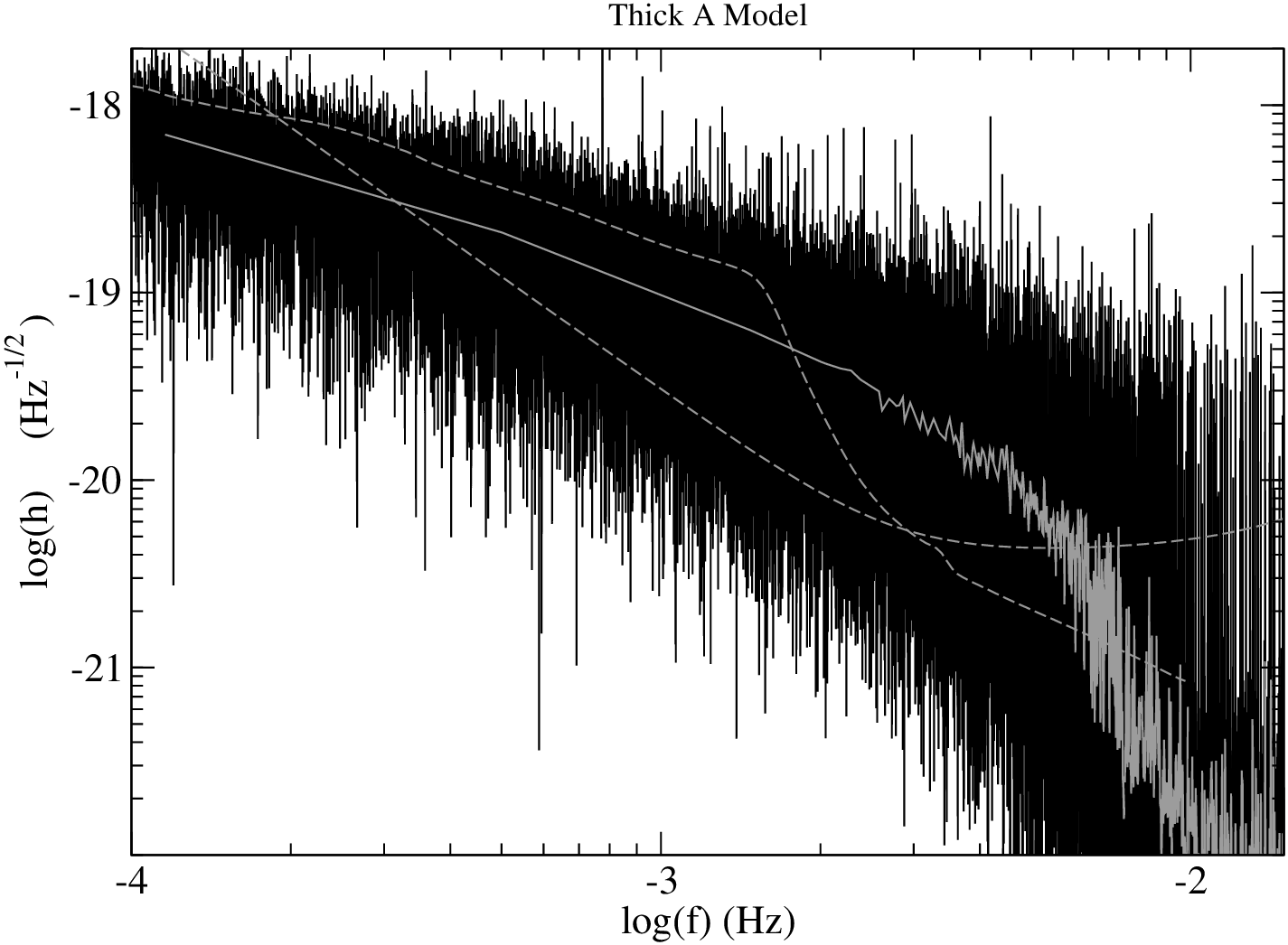}
\caption{Power Spectral Density (black) for the Thick A model with a running median over 1000 bins (gray). The expected noise level and standard~\citet{hils97} confusion limited noise from the LISA Sensitivity Curve Generator are shown as the dashed light gray curves. The noise and standard confusion limited noise curves have been multiplied by $\sqrt{3/20}$ to transform them from the barycenter frame to the detector frame.\label{thicka}}
\end{figure}

\begin{figure}
\plotone{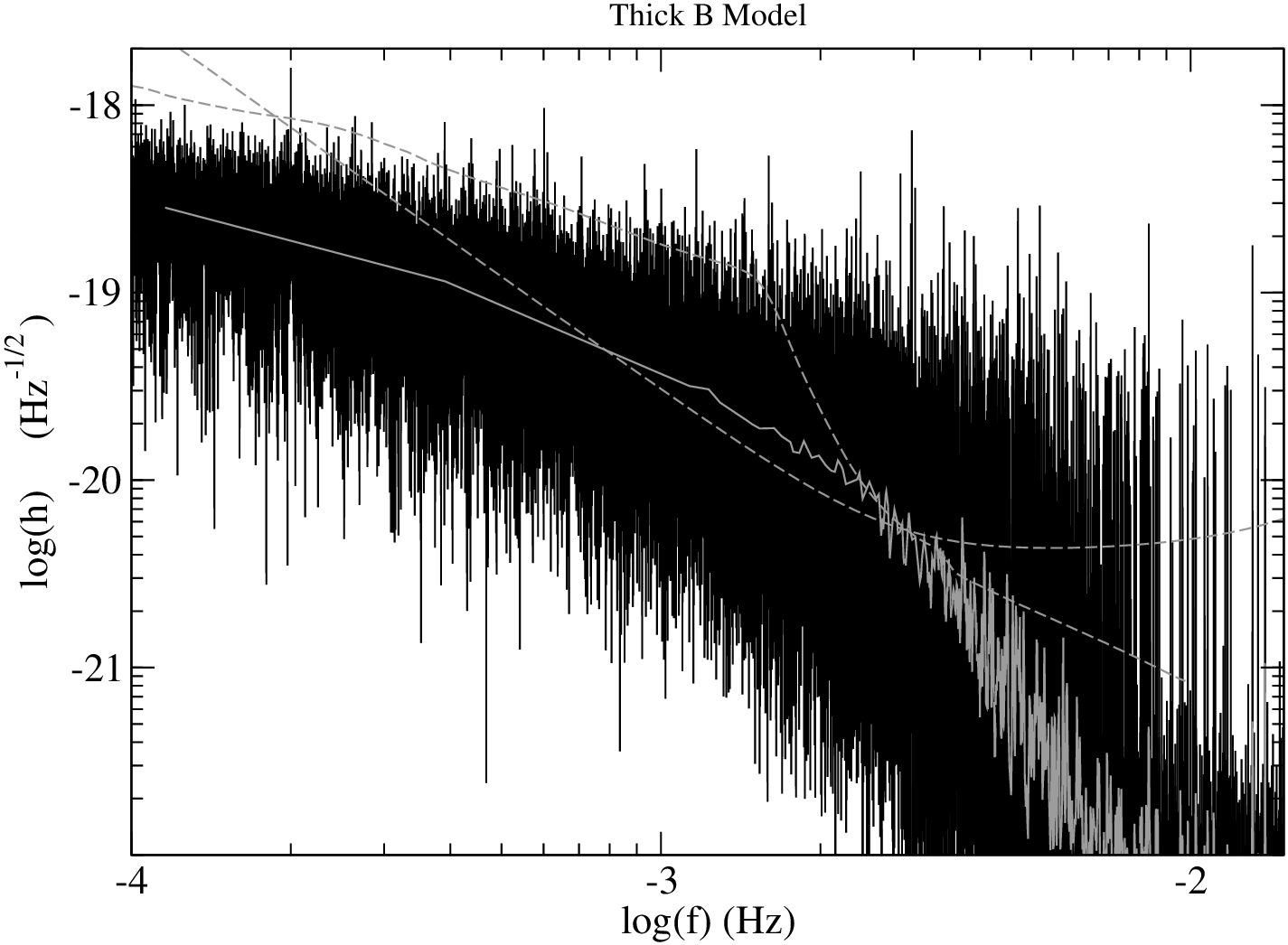}
\caption{Power Spectral Density (black) for the Thick B model with a running median over 1000 bins (gray). The expected noise level and standard~\citet{hils97} confusion limited noise from the LISA Sensitivity Curve Generator are shown as the dashed light gray curves. The noise and standard confusion limited noise curves have been multiplied by $\sqrt{3/20}$ to transform them from the barycenter frame to the detector frame.\label{thickb}}
\end{figure}

We have used the Michelson signal as the observable for the expected LISA data stream. The data stream is calculated in the long wavelength approximation for binaries whose central gravitational wave frequency is below 3 mHz (see~\citet{rubbo04} or~\citet{cutler98} for descriptions of this approximation). At higher frequencies ($f \gtrsim 3~{\rm mHz}$), we have used the rigid adiabatic approximation (see~\citet{rubbo04} or~\citet{vecchio04} for a description of this approximation). We have included a linear chirp for those binaries whose frequency will shift by more than $\sim 3 \times 10^{-9}$ Hz during the one year observation. We have generated 3 realizations of each of our Galaxy models. To minimize any variations between models, we used the same three initial random seeds to generate the three realizations of each model. Thus, the first realization of the Thin model has exactly the same population of binaries as the first realization of the Thick B model (modulo a rescaling of the $z$-coordinate), and the first realization of the Thick B model is a subset of the first realization of the Thick A model. Representative strain spectral densities for the three models are shown in Figures~\ref{thin}-\ref{thickb}. To better characterize the spectra, we have also included a running median over 1000 bins for each spectrum. For reference, we have also included the LISA sensitivity curve (at ${\rm SNR} = 1$) and the standard Hils-Bender CWDB confusion curve as generated by the {\it LISA Sensitivity Curve Generator}~\citep{larson03}. Because our spectra are calculated in the detector frame and the sensitivity curve generator output is in the barycenter frame, we have rescaled the sensitivity curve generator output by $\sqrt{3/20}$ to account for averaging over all polarizations and directions.
\clearpage

\begin{figure}
\plotone{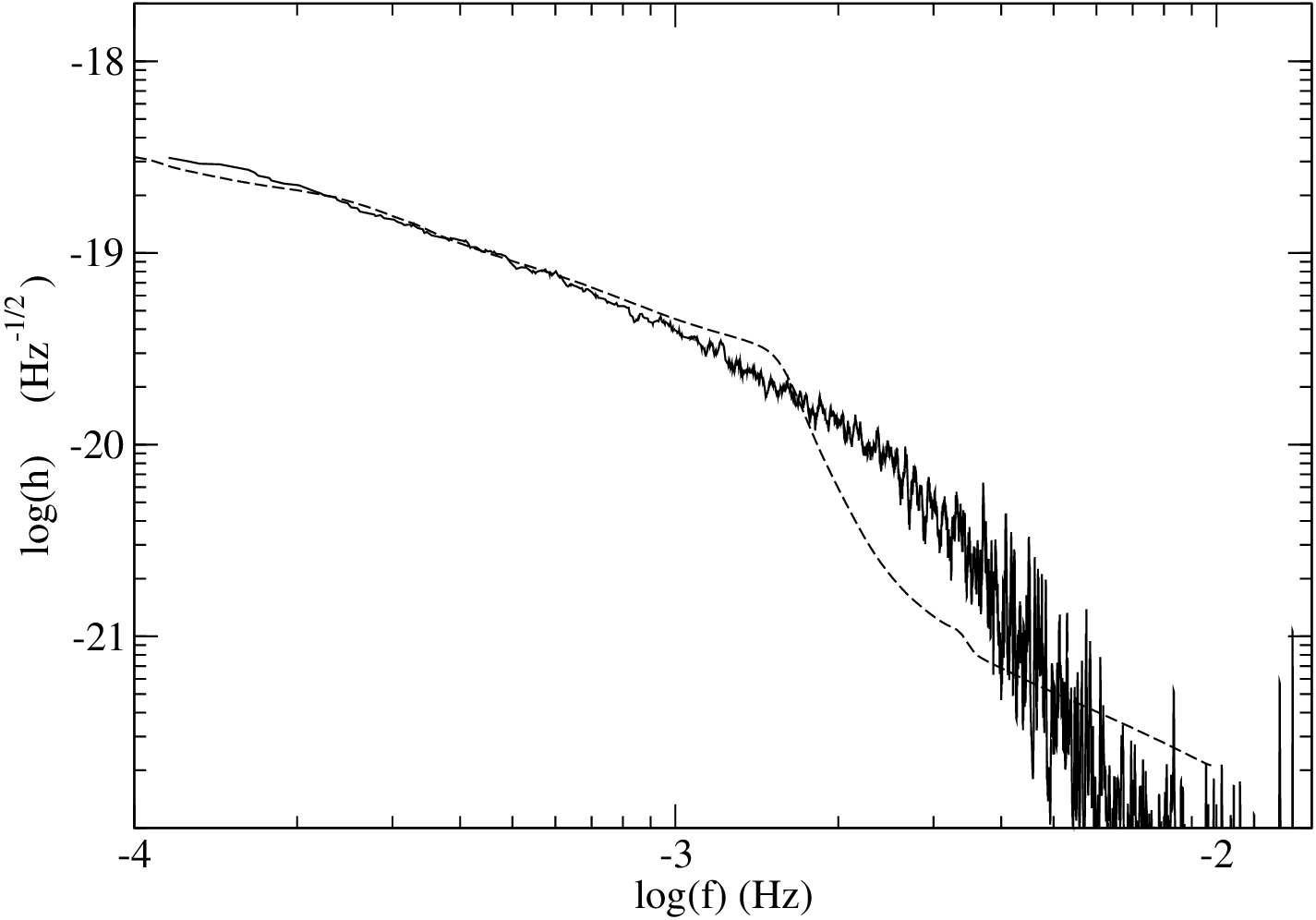}
\caption{A running median of the Thin model over $1000$ frequency bins compared with the standard CWDB confusion curve from the LISA Sensitivity Curve Generator scaled to coincide with the Thin model at low frequencies. \label{runmed}}
\end{figure}

\clearpage
The power spectra for the models show the expected behavior -- there is a foreground of strong nearby signals superimposed on a confusion-limited background at frequencies below about 1 mHz, numerous individually resolvable signals above about 10 mHz, and a transition zone in between where the floor of the power spectral density drops to the numerical noise level of $\sim 10^{-22}~{\rm Hz}^{-1/2}$.  We note that the running median of our spectra is generally below the standard Hils-Bender curve. This is due to the fact that we are using a population synthesis based upon the binary evolution of~\citet{nelemans01a}, and this tends to produce binaries whose chirp mass is roughly a factor of two lower than the population synthesis of~\citet{hbw90} (Nelemans, private communication). However, if one scales up the running median curve so that it coincides with the standard Hils-Bender curve one will notice that for each model there is still considerable signal noise beyond the standard upper confusion limit of 3 mHz, and that the noise level drops much more slowly than the standard confusion curve. This is shown in Figure~\ref{runmed} for the Thin model, which best matches the \citet{hbw90} CWDB population.  We caution, though, that it is misleading to quantitatively compare the specific characteristics of the standard CWDB curve to the Thin model -- the population synthesis model is different, the approach to data analysis is different, and we define a different measure of the confusion onset (see next section). 
Consequently, the standard confusion curve on these figures should be used merely as a qualitative benchmark.
\clearpage
\clearpage

\section{Estimating the Transition Frequency}\label{zetarationale}

Characterizing the shape and amplitude of the unresolvable gravitational wave sources is not a trivial undertaking, and many groups are working to develop techniques to extract both the resolvable signals from the confusion foreground (cf. \cite{umstatter05a,umstatter05b, cornish03a, cornish03b, takahashi02, cutler98}), and to quantify the extent of the foreground itself (\cite{hbw90, nelemans01b, bender97, evans87, postnov98, timpano05}). Since these techniques are still in development, there is no canonical and robust way to quantify the onset of the confusion limit. We develop and present one statistic in this section, though we caution that we are using it primarily as a comparison tool to discover trends {\em among our models}, and should not be used to compare our results with those of other published work. For a given data analysis technique, however, the same trends should be obtained.

The conversion of the raw galactic spectrum to a smooth curve representing the confusion limit is subject to a number of assumptions concerning both the ability to remove stronger foreground signals from the low-frequency region and to individually resolve and extract signals from the higher frequency regions where there are both full and empty frequency bins. Depending on how optimistically one chooses to estimate these abilities, one can dramatically alter the shape of the Galactic confusion curve. For example, the method described in \citet{hils97} assumes that information in three frequency bins are required to completely parameterize (and therefore subtract) the signal from an individual binary. These three bins are removed at the signal strength of the foreground binary. The effect of this assumption is to produce the dramatic drop in the standard white dwarf binary curve at around 3 mHz where the average number of binaries per frequency bin drops below 1. \citet{timpano05} have used a different method to estimate the ability to remove signals in the transition zone. Starting with the full signal from the entire population of binaries, they calculate a running median to determine the background level of the full signal. Next, they determine the bright binaries that stand above this level with a signal to noise of at least 5. These binaries are then competely and exactly removed from the data stream. The process is then repeated on the remaining signal until the number of new bright binaries is less than 1\% of the previously found bright binaries, resulting in a remaining signal that is nearly Gaussian. This process is highly optimistic and therefore represents a lower bound on the cleaned signal. \citet{nelemans01b,nelemans04} simply cut the signal off once the average number of binaries per bin drops below 1. Since these assumptions rely on the currently undemonstrated capabilities of future data analysis techniques, we choose to analyze the signal prior to any assumed data reduction. 

\subsection{The $\zeta$ Estimator}\label{zetasub}
Although individual binaries will occupy a single frequency bin in the barycenter frame (unless they are chirping), the motion of LISA will spread these signals out over several frequency bins. Consequently, simply counting the number of binaries per bin is not a very realistic way of determining the transition zone. The ability to resolve a signal is related more to the number of adjacent bins without signal rather than the number of binaries per bin. In order to successfully extract the signal due to a binary, it is necessary to determine parameters (such as sky position, orientation, and amplitude) that fully characterize the spreading of the signal in the frequency domain. The accuracy with which this can be done is related to the signal-to-noise ratio (SNR). Thus, we construct a parameter, $\zeta$ that characterizes the fraction of frequency bins that contain signal with a specified SNR. 

To motivate our definition of $\zeta$, we treat the signal strength $h(f)$ in any frequency bin as though it were sampled from a distribution function.
In frequency regions where the signal is completely confusion limited (or pure noise), we take this distribution function to be:
\begin{equation}
\label{zetastart}
P(h(f)) = \frac{h(f)}{\sigma_f^2}\exp{\left(\frac{-h^2(f)}{2\sigma_f^2}\right)}
\end{equation}
where $\sigma_f$ is a measure of the averaged signal strength over a suitably small frequency range so that it can be considered constant. From the mean:
\begin{equation}
\langle{h}\rangle = \int_{0}^{\infty}{hP(h)dh}
\end{equation}
and the variance:
\begin{equation}
\sigma_h^2 = \int_{0}^{\infty}{h^2P(h)dh} - \langle{h}\rangle^2
\end{equation}
we can construct the dimensionless quantity:
\begin{equation}
\label{zeta}
\zeta = \frac{\sqrt{\sigma_h^2}}{\langle{h}\rangle} = \sqrt{\frac{4-\pi}{\pi}}.
\end{equation}
Thus, in the confusion limited regime, $\zeta$ should be well described by Equation~\ref{zeta}, and should be approximately 0.5.

When the data stream consists of both noise and signal, the signal strength in any frequency bin should then be modeled as if it were sampled from a linear combination of two distribution functions, one for the noise and one for the signals. This function is given by:
\begin{equation}
P^{\prime}(h(f))  =   a\frac{h(f)}{\sigma_f^2}\exp{\left(\frac{-h^2(f)}{2\sigma_f^2}\right)}  
 +  (1 - a)\frac{h(f)}{(b\sigma_f)^2}\exp{\left(\frac{-h^2(f)}{2(b\sigma_f)^2}\right)},
\end{equation}
where $a$ gives the probability that a given frequency bin will be empty (i.e.: dominated by instrument noise of strength $\sigma_f$) and $b$ is a measure of the signal-to-noise ratio for the average resolvable signal. The value of $\zeta$ in this case is given by:
\begin{equation}
\label{criterion}
\zeta = \frac{\left[4\left(a + b^2\left(1-a\right)\right) - \pi\left(a+b\left(1-a\right)\right)^2\right]^{1/2}}{\sqrt{\pi}\left(a+b\left(1-a\right)\right)}.
\end{equation}

When we apply $\zeta$ to our simulation of the LISA data stream, we compute the mean and variance of the strain spectral density over $n$ frequency bins using:
\begin{eqnarray}
\label{mean}
\langle{h}\rangle_i = \frac{1}{n}\sum_{j=i}^{i+n-1}{h\left(f_j\right)} \\
\label{var}
\sigma_{hi}^2 = \frac{1}{n}\sum_{j=i}^{i+n-1}{h^2\left(f_j\right)}\end{eqnarray}
and construct:
\begin{equation}
\zeta_i = \frac{\sqrt{\sigma_{hi}}}{\langle{h_i}\rangle}.
\end{equation}
In a region of the spectrum where the potentially resolvable sources have SNR given by $b$, then $\zeta_i$ is a measure of the fraction of bins devoid of resolvable signal. Consequently, one can set a threshold value of $\zeta_i$ corresponding to the value of $a$ necessary for a given data analysis technique to be able to resolve signals with SNR $\gtrsim b$.

\subsection{Expected $N$-dependence of Transition Frequency}\label{ntestsub}
Without adhering to a specific data analysis technique, the choice for the threshold value of $\zeta_i$ is somewhat arbitrary. However, for a given choice of $a$, we can expect the transition frequency ($f_t$) at which $\zeta_i$ crosses the associated threshold to scale with the total number of binaries ($N$) according to $f_t \propto N^{3/8}$. In order to understand this scaling, we note that in the expected transition region between 1 and 10 mHz, the number density of binaries in frequency space is governed primarily by the period evolution through gravitational radiation. Therefore, according to~\citet{hbw90}
\begin{equation}
\frac{dN}{df} \propto Nf^{-11/3},
\end{equation}
where $f$ is the gravitational wave frequency. At the frequencies of the transition region, the primary source of spreading of the signal into adjacent frequency bins is the Doppler modulation of the signal, so $\Delta f \propto f$. Consequently, at the transition frequency
\begin{equation}
\left.\frac{dN}{df}\right|_{f_t} \propto f_t^{-1} \propto N f_t^{-11/3},
\end{equation}
and so we obtain $f_t \propto N^{3/8}$.

In order to test the ability of the $\zeta$ estimator to correctly reproduce the expected scaling of $f_t$ with $N$, we have applied $\zeta$ to a broad set of Galaxy realizations using the density distribution given by Eq.~\ref{cuspyrho} with different values for $z_0$, $R_0$ and $N$. Note that these toy models have a different density distribution than our disk population models, which more closely match~\citet{hbw90}; these models serve to test the $\zeta$ estimator and determine reasonable values for $a$, $b$, and $\zeta_i$. In these realizations, we did not introduce any simulated instrumental noise into the data stream, but numerical round-off errors introduce an effective noise contribution that we estimate from the strain spectral density to be $\simeq 10^{-22}~{\rm Hz}^{-1/2}$ In the expected transition region the mean signal strength, read from the strain spectral density is $\simeq 10^{-20}~{\rm Hz}^{-1/2}$. Correspondingly, we take $b = 100$ as an approximation of the SNR in the region of interest. We arbitrarily chose $a = 0.5$ to give a threshold value of $\zeta_i = 1.2$. We then chose to average over $n = 20,000$ bins to ensure a reasonably smooth curve for $\zeta$. If smaller values of $n$ are used, the effect is to increase the variation in $\zeta$, which increases the spread of frequencies for which $\zeta$ passes through the threshold without significantly altering the central value of $f_t$. The values of $f_t$ obtained in this way are shown in Table~\ref{ntest}. A least-squares fit to the values of $f_t$ for a function of the form $f_t = \alpha N^{\beta}$ gives $\beta = 0.39 \pm 0.04$ indicating that $\zeta$ does reproduce the correct scaling of the transition frequency with total number.
\clearpage
\begin{deluxetable}{rcl}
\tablecaption{Transition frequencies at which $\zeta$ passes through 1.2 for Galaxy realizations with different $N$.\label{ntest}}
\tablehead{\colhead{$N$ ($\times 10^6$)} & \colhead{Number of Models} & \colhead{$f_t$ (mHz)}}
\startdata
4 & 3 & $2.90 \pm 0.07$ \\
8 & 6 & $4.01 \pm 0.11$ \\
12 & 1 & $4.64$ \\
20 & 2 & $5.56 \pm 0.18$ \\
\enddata
\end{deluxetable}
\clearpage

We note that the $\zeta$ estimator is really nothing more than a dimensionless measure of the degree of scatter in the strain amplitude about its mean. It is through the motivation described in Eqs~\ref{zetastart}-\ref{criterion} that we interpret $\zeta$ as a measure of the number of resolvable signals with SNR $> b$. The fact that $\zeta$ reproduces the expected scaling of the transition frequency with total number can be viewed as confirmation that this interpretation is reasonable.

\section{Results}
\begin{figure}
\plotone{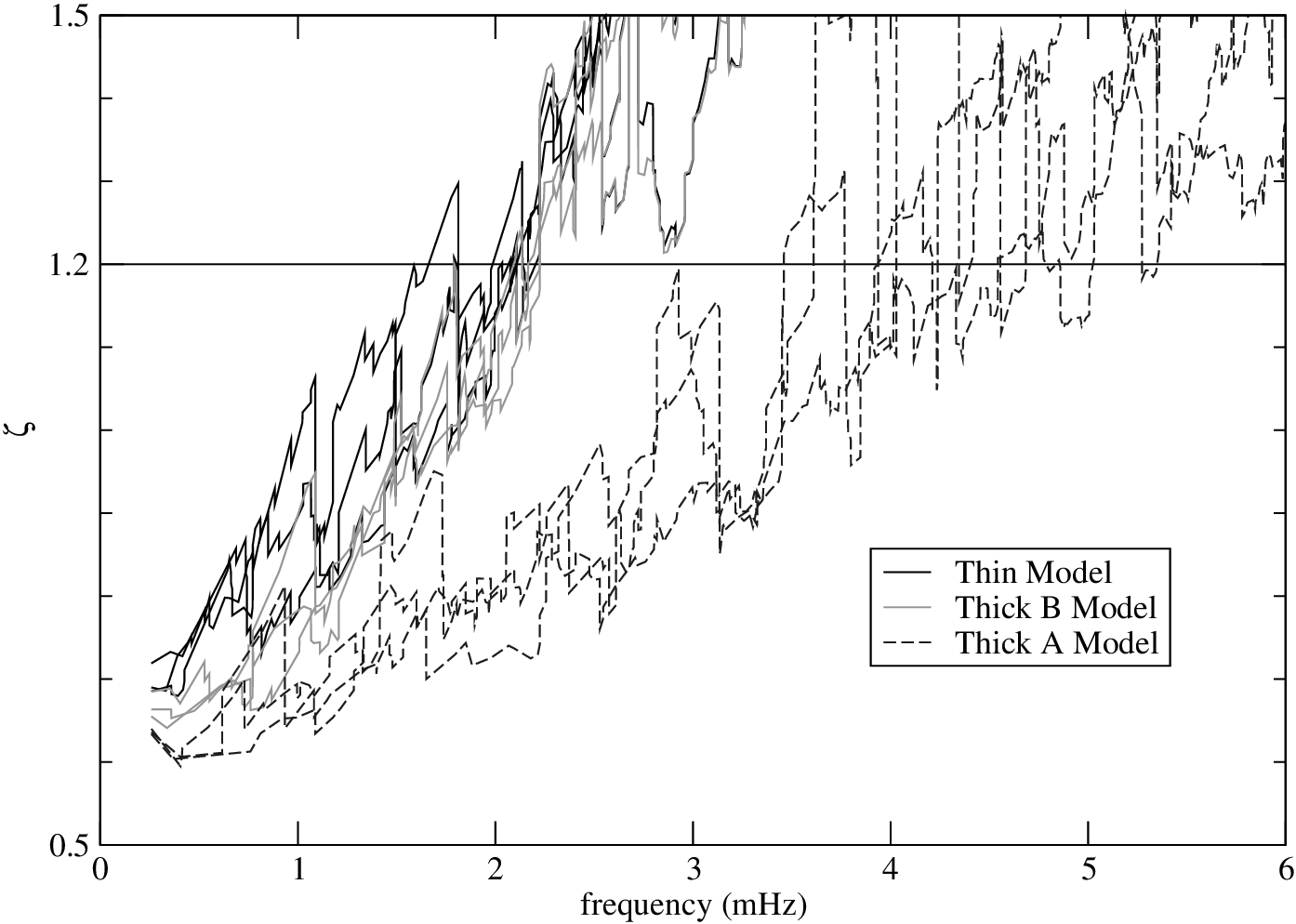}
\caption{$\zeta$ for all models in the transition zone. Three realizations for each model are plotted with Thin (black), Thick A (dashed dark gray), and Thick B (light gray). \label{zetaftplots}}
\end{figure}

We have applied the $\zeta$ estimator to our model populations to determine the relationship between the transition frequency and the scale height. We calculated the mean and variance as described in Eqs.~\ref{mean},~\ref{var}, averaging over $n = 20,000$. Again, as in section~\ref{ntestsub}, by inspection of the strain spectral density plots, we find that the mean signal strength in the expected transition zone is $\sim 10^{-20}$, while the effective noise due to numerical round-off errors is $\sim 10^{-22}$. Consequently, we take the same values as in section 3.2. We note that $\zeta$ is weakly dependent upon the value of $b$, so that reducing $b$ by a factor of 5 (as might naively be expected from increasing the scale height by the same factor) results in a negligible reduction of the threshold value to $\zeta = 1.15$. The plots of $\zeta$ for all models are shown in Figure~\ref{zetaftplots}.

\clearpage
\clearpage

\begin{figure}
\plotone{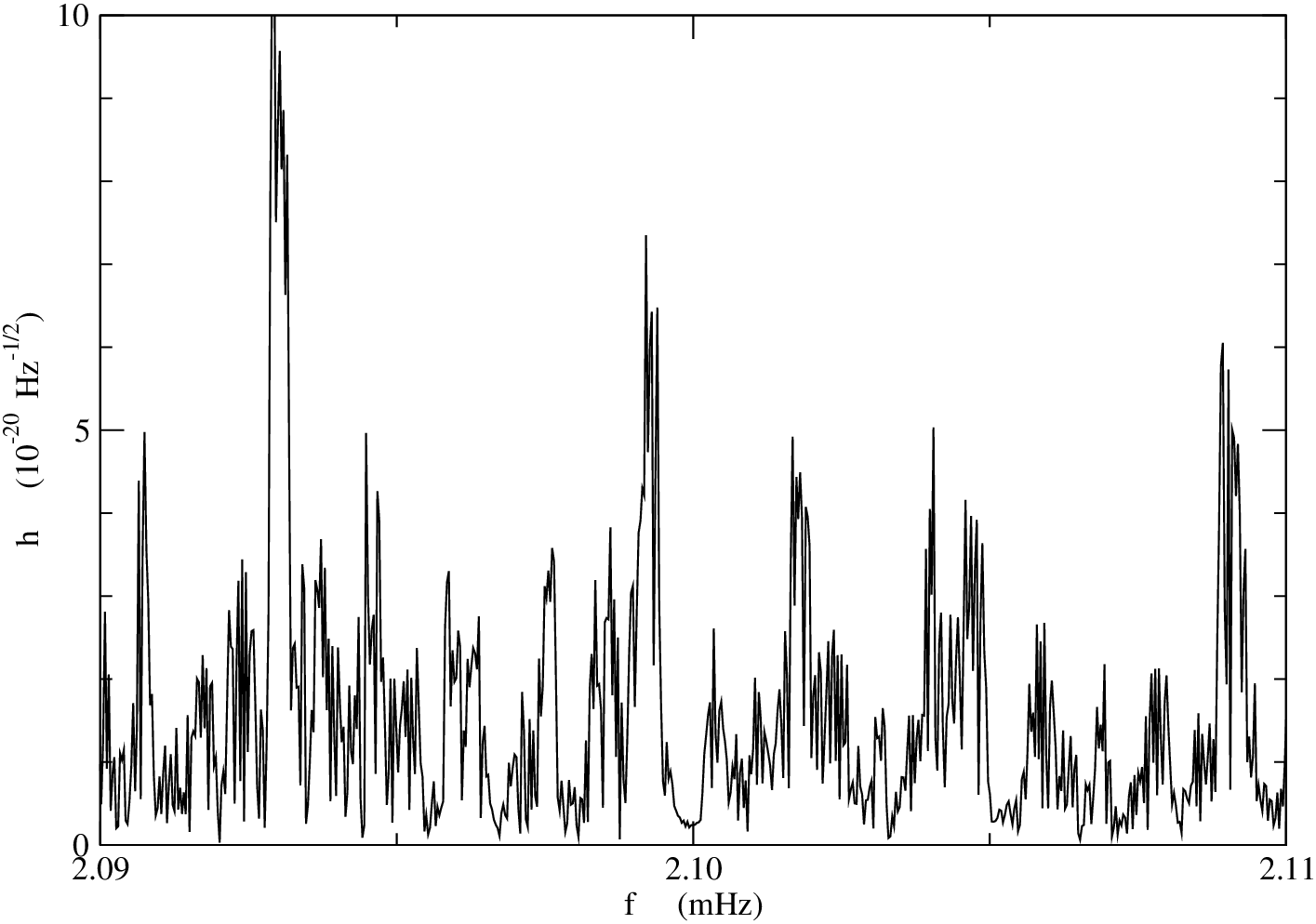}
\caption{Strain spectral density for the Thin model between 2.09 and 2.11 mHz where $\zeta$ passes through 1.2. There are 76 binaries in this region of the spectrum. \label{transition}}
\end{figure}

Because the transition between confusion-limited signal to individually resolvable sources occurs over a spread of frequencies, $\zeta$ can fluctuate about 1.2 over a range of frequencies. Furthermore, regions where random clustering of particularly loud signals can stand well above the confusion-limited noise will also produce values of $\zeta$ above 1.2. The range of frequencies through which $\zeta$ transitions past 1.2 are given for each model in Table~\ref{crittable}. It is clear from Figure~\ref{zetaftplots} that using a larger scale height for the Galaxy while maintaining the local space density of binaries results in a higher transition frequency, as would be expected from simply increasing the total number of binaries in the Galaxy. Also, we note that there is no significant difference in the value of $f_t$ between the Thin and Thick B models. Again, this is not unexpected since all of these models contain the same number of binaries. We show in Figure~\ref{transition} the appearance of the spectrum for the Thin model around 2.1 mHz to illustrate the separation of signals at this point. There are 76 binaries shown in Figure~\ref{transition}.

\clearpage
\begin{deluxetable}{lcr}
\tablecaption{Transition frequencies for the disk population models.\label{crittable}}
\tablehead{\colhead{Model} & \colhead{Realization} & \colhead{$f_t$ (mHz)}}
\startdata
Thin & 1 & 2.09 - 2.11 \\
Thin & 2 & 1.79 - 2.17\\
Thin & 3 & 1.58 - 2.20\\
Thick A & 1 & 2.93 - 4.42\\
Thick A & 2 & 3.60 - 5.03\\
Thick A & 3 & 3.46 - 5.35\\
Thick B & 1 & 2.22\\
Thick B & 2 & 1.79 - 2.19\\
Thick B & 3 & 2.10 - 2.23
\enddata
\end{deluxetable}
\clearpage

\clearpage

\begin{figure}
\plotone{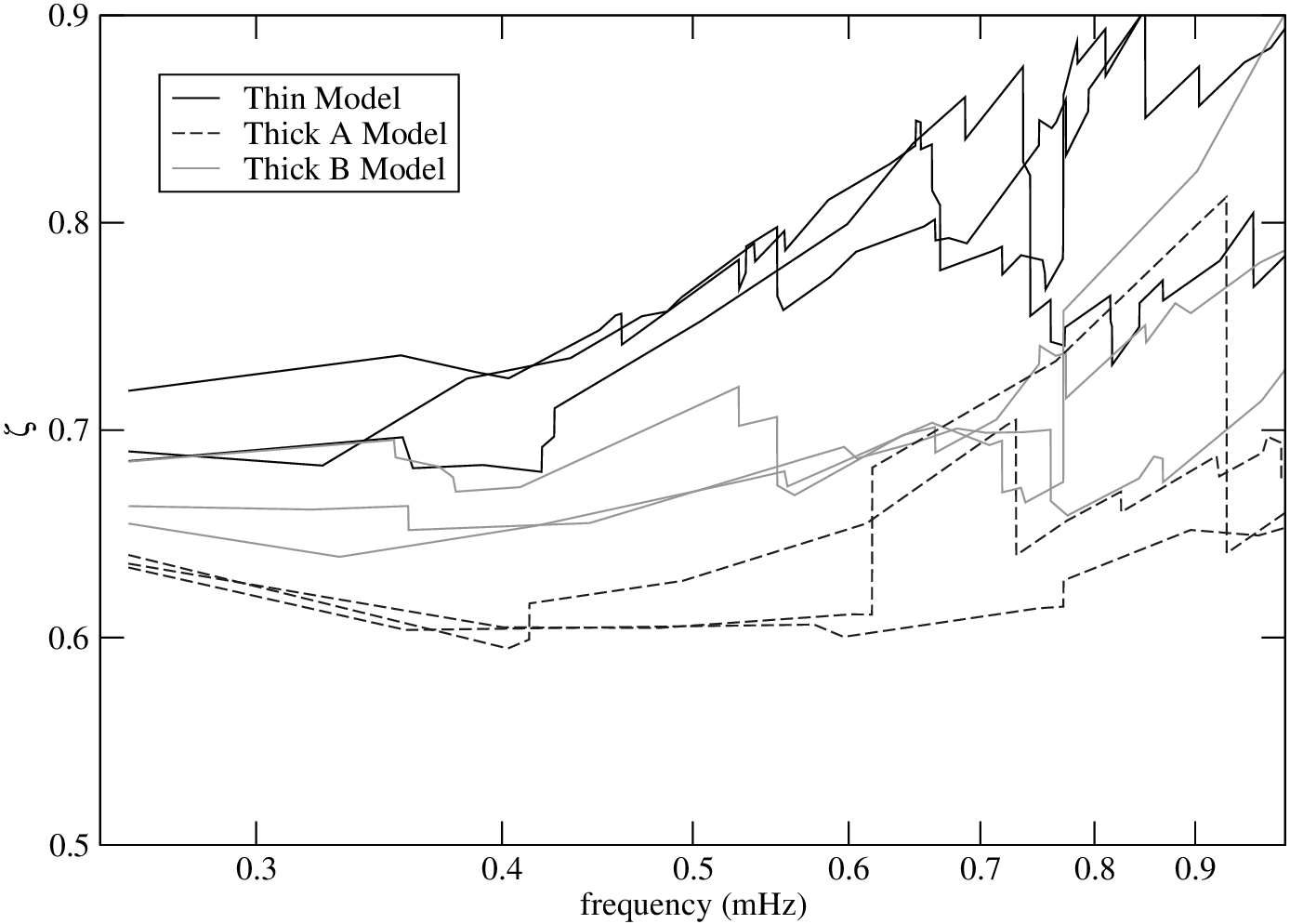}
\caption{$\zeta$ as a function of frequency (in mHz) for the three realizations of each model. Note that Thin (black) is above Thick B (light gray) which is above Thick A (dashed dark gray).\label{lowfzetaplots}}
\end{figure}

At the low frequency end of Figure~\ref{zetaftplots}, there is an interesting distinction between all three models. Below about 1 mHz, the Thin model has a higher value of $\zeta$ than either Thick model, and the Thick A model has a lower value of $\zeta$ than the other two models. This detail is shown in Figure~\ref{lowfzetaplots}. It is understandable that the Thick A realizations should should give a $\zeta$ below the Thin and Thick B models. The much larger total number of binaries in the Thick A models will produce a stronger confusion-limited signal from the thousands of binaries per frequency bin expected in the low frequency end of the spectrum. Consequently, we would expect an overall smaller SNR for the bright signals that stand out above the background. This, in turn will produce a smaller value of $\zeta$ due to the reduced value of $b$.

\clearpage
\clearpage

The explanation for the differences between the Thin and Thick B models is a little more subtle. First, we note that the background confusion-limited signal should be similar for both models, since they both have the same value of $N$ (and therefore the same number of binaries per frequency bin). Furthermore, we used the same three random seeds to generate the three realizations of the Thin model as we used to generate the three realizations of the Thick B model, so the number density of binaries in frequency space cannot differ between the two models. Therefore, the differences in $\zeta$ between these two models cannot arise from different values of $a$. However, since the Thick B models have a larger scale height, the local space density is reduced by the ratio of scale heights. This, in turn, means that the average distance to the nearby binaries is larger in the Thick B models compared with the Thin models. Again, this will result in an overall lowering of the SNR of the bright signals compared with the Thin models, producing a smaller value of $\zeta$ in the Thick B models.

\section{Conclusions}\label{conclusions}

The canonical curve that is used to represent the Galactic white dwarf binary confusion-limited signal in descriptions of LISA's sensitivity is based upon a {\it de facto} local space density calibration of the total number of binaries in the Galaxy. The spatial distribution model that was used has a scale height of 90 pc~\citep{hbw90}. More recent models of the Galactic white dwarf binary population use a spatial distribution model with a scale height of 200 pc~\citep{nelemans01a,nelemans01b}. Current estimates for the vertical scale height for white dwarfs in the disk are somewhat thicker with scale heights between 240 and 500 pc~\citep{nelson02}. The consequences of using an increased scale height in models of the Galactic white dwarf binary contribution to the LISA sensitivity curve differ depending upon whether one calibrates the total number of binaries using global properties (such as inferred star formation history) or local properties (such as local space density). We have generated three models of the Galactic white dwarf binary population and calculated the expected LISA data stream from three realizations of each of these models. In order to analyze the expected transition frequency of the signal as it goes from confusion-limited to individually resolvable, we have introduced an estimator, $\zeta$.

If the total number of binaries ($N$) in a Galactic binary simulation is calibrated by using the local space density, then an increase in the scale height of the spatial distribution model will result in a linear increase in the total number of binaries. In this case, the resulting change to the expected confusion curve is what one would expect from simply increasing the number of binaries. The transition frequency increases and the overall level of the confusion-limited signal at low frequencies is increased. Since the local space density is used to calibrate $N$, the number of nearby low frequency sources is unchanged by going to a larger scale height. The result of this is that fewer of these nearby sources will stand out above the low frequency confusion-limited signal. Consequently, successfully extracting these bright sources will have less of an effect on the low frequency confusion-limited signal than might be assumed from using a thin disk model, such as found in~\citet{timpano05}.

If the total number of binaries is fixed by a global calibration method, the resulting effect of changing the scale height is considerably more subtle. Since the value of $N$ is independent of the scale height, the expected transition frequency will not change. This is borne out by our analysis with the $\zeta$ estimator. In addition, the overall level of the low frequency confusion-limited signal will change only negligibly due to the slight increase in the average distance to distant binaries. However, there can still be an effect on the low frequency confusion-limited signal. With a fixed $N$, the local space density depends inversely on the scale height so a larger scale height yields fewer expected nearby bright sources. Furthermore, the average distance to these nearby sources will increase for models with greater scale height. The net result is that fewer bright sources will be extractable from the low frequency confusion-limited signal.
 
\acknowledgements
We would like to thank the anonymous referee for insightful comments which have helped us achieve a clearer picture of the effect of scale height on the confusion limit. We would like to acknowledge the support of the Aspen Center for Physics and the Center for Gravitational Wave Physics, which is funded by the National Science Foundation under the cooperative agreement PHY 01-14375. MJB also acknowledges the support of NASA APRA grant no. NNG04GD52G. Work at the Aspen Center was supported by NASA Award Number NNG05G106G.

\end{document}